\documentclass[10pt,iop]{emulateapj} 
\usepackage{ctable}
\usepackage{amsmath}
\pdfoutput=1

\begin{document}

\submitted{Accepted by AJ: 12 Dec 2011}
\title{A Technique for Primary Beam Calibration of Drift-Scanning, Wide-Field Antenna Elements}
\author{Jonathan C. Pober\altaffilmark{1}, Aaron R. Parsons\altaffilmark{1}, Daniel C. Jacobs\altaffilmark{2}, James E. Aguirre\altaffilmark{3}, Richard F. Bradley\altaffilmark{4,5,6}, Chris L. Carilli\altaffilmark{7}, Nicole E. Gugliucci\altaffilmark{6}, David F. Moore\altaffilmark{3}, Chaitali R. Parashare\altaffilmark{4}}

\altaffiltext{1}{Astronomy Dept., U. California, Berkeley, CA}
\altaffiltext{2}{School of Earth and Space Exploration, Arizona State U., Tempe, AZ}
\altaffiltext{3}{Dept. of Physics and Astronomy, U. Pennsylvania, Philadelphia, PA}
\altaffiltext{4}{Dept. of Electrical and Computer Engineering, U. Virginia, Charlottesville, VA}
\altaffiltext{5}{National Radio Astronomy Obs., Charlottesville, VA}
\altaffiltext{6}{Dept. of Astronomy, U. Virginia, Charlottesville, VA}
\altaffiltext{7}{National Radio Astronomy Obs., Socorro, NM}

\begin{abstract}
We present a new technique for calibrating the primary beam of a wide-field,
drift-scanning antenna element.  Drift-scan observing is not compatible with standard
beam calibration routines, and the situation is further complicated by
difficult-to-parametrize beam shapes and, at low frequencies, the sparsity of accurate
source spectra to use as calibrators.  We overcome these
challenges by building up an interrelated network of source ``crossing points" ---
locations where the primary beam is sampled by multiple sources.  Using
the single assumption that a beam has 180$^\circ$ rotational symmetry, we can
achieve significant beam coverage with only a few tens of sources.
The resulting network of crossing points
allows us to solve for both a beam model and source flux densities referenced
to a single calibrator source, circumventing the need for a large sample of
well-characterized calibrators.  We illustrate the method with actual and
simulated observations from the Precision Array for Probing the Epoch of
Reionization (PAPER).
\end{abstract}


\section{Introduction}

The past decade has seen a renewed interest in low frequency radio astronomy with a strong focus on cosmology with the highly redshifted 21cm line of neutral hydrogen.  
Numerous
facilities and experiments are already online or under construction, including the Giant Metre-Wave Radio Telescope (GMRT; \citealt{gmrt})\footnote{http://gmrt.ncra.tifr.res.in/}, the LOw Frequency
ARray (LOFAR; \citealt{lofar})\footnote{http://www.lofar.org/}, the Long Wavelength Array (LWA; \citealt{lwa})\footnote{http://www.phys.unm.edu/~lwa/index.html} and the associated Large Aperture-experiment to Detect the Dark Ages (LEDA) experiment, the Cylindrical Radio Telescope (CRT/BAORadio, formerly HSHS, \citealt{crt1,crt2})\footnote{http://cmb.physics.wisc.edu/people/lewis/webpage/index.html},
the Experiment to Detect the Global EoR Step (EDGES; \citealt{edges})\footnote{http://www.haystack.mit.edu/ast/arrays/Edges/}, 
the Murchison
Widefield Array (MWA; \citealt{mwa})\footnote{http://www.mwatelescope.org/}, 
and the Donald C. Backer
Precision Array for Probing the Epoch of Reionization
(PAPER; \citealt{parsons_et_al2010})\footnote{http://eor.berkeley.edu/}.  21cm cosmology experiments will need to
separate bright galactic and extragalactic foregrounds from the neutral hydrogen signal,
which can be fainter by as much as 5 orders of magnitude or more (see, e.g.,
\citealt{furlanetto_et_al2006} and \citealt{santos_et_al2005}).  As such, an
unprecedented level of instrumental calibration will be necessary for the detection
and characterization of the 21cm signal.

Achieving this level of calibration accuracy is complicated by the design
choice of many experiments to employ non-tracking antenna elements (e.g.
LWA, MWA, LOFAR and PAPER).  Non-tracking elements can provide significant reductions in cost compared to traditional dishes, while also offering increased system stability and smooth beam responses.  
However, non-tracking elements also present many calibration
challenges beyond those of traditional radio telescope dishes.  Most
prominently, the usual approach towards primary beam calibration --- pointing
at and dithering across a well-characterized calibrator source --- is not
possible.  Instead, each calibrator can only be used to characterize the small
portion of the primary beam it traces out as it passes overhead.
Additionally, the wide fields of view of many elements make it non-trivial to
extract individual calibrator sources from the data (see, e.g., \citealt{parsons_and_backer2009} and \citealt{paciga_et_al2011} for approaches to isolate calibrator sources).  Finally, many of these
arrays use dipole and tile elements, the response of which are not easily described by simple analytic functions.

In this paper, we present a method for calibrating the primary beams of
non-tracking, wide-field antenna elements using astronomical sources.  We illustrate the
technique using both simulated and observed PAPER data from a 12 antenna array
at the NRAO site in Green Bank, WV.  PAPER is an interferometer operating 
between 100 and 200 MHz, targeted towards
the highly-redshifted 21cm signal from the epoch of reionization.  Although
the cosmological signal comes from every direction on the sky, an accurate primary
beam model will be necessary to separate the faint signal from bright foreground emission.
In this work, we use a subset of the brightest extragalactic sources to calibrate the primary beam.
Because interferometers like PAPER are insensitive to smooth emission
on large scales, these extragalactic sources are easily detectable despite the
strongly increasing brightness of Galactic
synchrotron emission at low radio frequencies.  Only the measured relative flux densities of each 
source are needed to create a beam model, but to facilitate comparision with other catalogs,
we use the absolute spectrum of Cygnus A from \citet{baars_et_al1977} to place our source
measurements on an absolute scale.

The structure of this paper is as follows:
in \S\ref{motivation} we motivate the problem and the need for a new approach to primary
beam calibration for wide-field, drift-scanning elements.
In \S\ref{methods} we present our technique for primary beam calibration.  We
show the results of applying the method to simulated and actual observations in
\S\ref{sims} and \S\ref{data}, respectively, and we conclude in
\S\ref{conclusions}.

\section{Motivation}
\label{motivation}

For non-tracking arrays with static pointings such as PAPER, every celestial source
traces out a repeated ``source track" across the sky, and across the beam,
each sidereal day. The basic relationship between the perceived
source flux density (which we shall call a measurement, $M$) measured at time $t$ and the intrinsic
source flux density ($f$) is:
\begin{equation}
\label{basicmeas}
M(t) = b(\hat s(t))f, %
\end{equation}
where $b$ is the response of the primary beam toward the time-dependent
source location $\hat s$.

If the inherent flux density of each source were well-known, it would be
straightforward to divide each $M(t)$ by $f$ to obtain
$b$ along the source track $\hat s(t)$.  To form a complete beam model, one would then need enough well-characterized sources to cover the entire sky.  In the 100-200 MHz band, however, catalog accuracy
for most sources is lagging behind the need for precise beam calibration
\citep{vollmer_et_al2005,jacobs_et_al2011}, which in turn is necessary
for generating improved catalogs.  
Without accurate source flux densities, both $b$ and
$f$ are unknowns, and Equation \ref{basicmeas} is underconstrained.  The problem
becomes tractable if several sources pass through the same location on the sky,
and therefore are attenuated by the same primary beam response.  However, the density of bright sources at low frequencies is insufficient to relate sources at different declinations.  
Additional information is necessary to break the degeneracy between
primary beam response and the inherent flux densities of sources.

\section{Methods}
\label{methods}

\begin{figure}
\centering
\includegraphics[scale=.4]{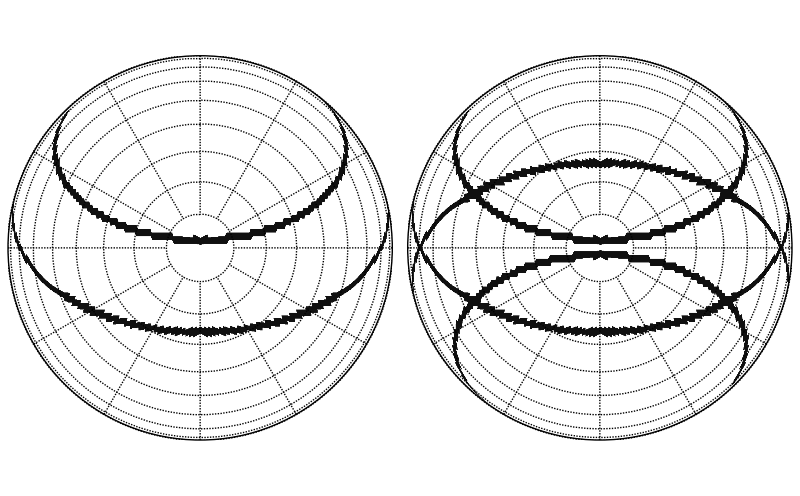}
\caption{
The building up of source tracks and crossing points.  \emph{Left:} The paths
Cygnus A (top) and Virgo A (bottom) take across the beam for the PAPER array in Green Bank.
The projection is orthographic with zenith at the center; dotted lines are
$10^{\circ}$ and $30^{\circ}$ steps in elevation and azimuth, respectively.
\emph{Right:} The tracks of Cygnus A and Virgo A, and their $180^{\circ}$
rotated images overlaid.  There are now 6 crossing points.  These are the
locations where the beam and source parameters can be solved for independently.
}
\label{trackexample}
\end{figure}

One way to overcome the beam-response/flux-density degeneracy described above is to assume
$180^{\circ}$ rotational symmetry in the beam.  Under this assumption, each
source creates two source tracks across the sky: one corresponding to the
actual position of the source, and the other mirrored across the beam center.
Under this symmetry assumption, tracks overlap at ``crossing points," as schematically
illustrated in Figure \ref{trackexample} for two sources: Cygnus A and Virgo
A.  
At a crossing point, there are only 3 unknowns, since each source
is illuminated by the same primary beam response, $b$.  Since two circles on the sky
cross twice (if they cross at all), there will be two independent relations that 
together provide enough information to constrain both the primary beam response
at those points and the flux densities of sources passing
through them (relative to an absolute flux scale).  
By using multiple sources, it is possible to build a network of
crossing points that covers a large fraction of the primary beam.  Furthermore,
such a network allows data to be calibrated to one well-measured fiducial
calibrator.  For observations with the Green Bank deployment of PAPER, 
source fluxes are related to Cygnus A from \citet{baars_et_al1977}.

In the rest of this section, we discuss the algorithmic details of
how this approach to primary beam calibration is implemented.  These steps
are:
\begin{enumerate}
\item extracting measurements of perceived source flux densities versus time from observations
(\S\ref{pfluxes}),
\item entering measurements into a gridded sky and finding
crossing points (\S\ref{gridding}),
\item solving a least-squares
matrix formulation of the problem (\S\ref{lstsqs}), and, finally,
\item deconvolving the
irregularly sampled sky to create a smooth beam (\S\ref{deconvolution}).  
\end{enumerate}
We also discuss prior information that can be included to refine the beam model
in \S\ref{priors}.  

\subsection{Obtaining Perceived Source Flux Densities}
\label{pfluxes}

The principal data needed for this approach are measurements of perceived flux
density versus time for multiple calibrator sources as they drift through the
primary beam.   In practice, any method of extracting perceived individual
source flux densities (such as image plane analysis) could be used; the beam
calibration procedure is agnostic as to the source of these measurements.  In this
work, we use delay/delay-rate (DDR) filters to extract estimates of individual
source flux densities as a function of time and frequency.  The frequency information can be used to perform a frequency-dependent beam calibration if the SNR in the observations is high enough.  These filters work in
delay/delay-rate space (the Fourier dual of frequency/time space) to isolate
flux density per baseline from a specific geometric area on the sky; for a full description
of the technique, the reader is referred to \citet{parsons_and_backer2009}.  

The first step of our pipeline to produce perceived flux density estimates is
to filter the Sun from each baseline individually in DDR space.  This is done
to ensure that little to no signal from the Sun, which is a partially-resolved
and time-variable source, remains in the data.  
(In principle, data from different observing runs separated by several months could
provide complete sky coverage while avoiding daytime data altogether.  We chose to use
data from only one 24 hour period to minimize the effect of any long
timescale instabilities in the system).  
A Markov-Chain Monte Carlo
(MCMC) method for extracting accurate time- and frequency-dependent source
models via self-calibration using DDR-based estimators is then used to 
model the 4 brightest sources remaining: Cygnus A, Cassiopeia A, Virgo A,
and the Crab Nebula (Taurus A).  The MCMC aspect of this algorithm iterates on a simultaneous model of all sources being fit for to minimize sidelobe cross-terms between sources.
After removing the models of Cygnus A, Cassiopeia A, Virgo A and
Taurus A from the data, a second pass of the MCMC DDR self-calibration algorithm extracts models of the
remaining 22 sources listed in Table \ref{srctable}.

\subsection{Gridding the Measurements}
\label{gridding}

We can increase the signal-to-noise at a crossing point by combining all
measurements within a region over which the primary beam response can be assumed to be constant.  To define these regions, we grid the sky.
In this work, the beam model is constructed on an
HEALPix map \citep{gorski_et_al2005} with pixels $\approx0.9^{\circ}$ on a side.  The choice of grid pixel size
is somewhat arbitrary.  Using a larger pixel size broadens the beam coverage of 
each source track, creating more crossing points and helping to constrain the
overall beam shape.
However, when the pixels are too large, each pixel includes
data from sources with larger separations on the sky.
Since the principal tenet of this approach is that each source within a crossing
point sees the same primary beam response, excessively 
large pixels can violate this assumption, resulting in an inaccurate beam model.
For PAPER data, a HEALPix grid with $0.9^{\circ}$ pixels is found to
be a good balance between these competing factors, as will be explained in
\S\ref{beamsims}.  For other experiments with narrower, more rapidly evolving
primary beams, smaller pixels may be necessary.

To introduce our measurements of perceived flux density into the grid, we first
recast Equation \ref{basicmeas} into a discrete form:
\begin{equation}
\label{healmeas}
M_i = b_if_k,
\end{equation}
where $M_i$ and $b_i$ are the respective perceived source flux densities and primary beam
responses in the pixel $i$, and $k$ is a source index labelling the inherent source flux density, $f$.  To generate a
single measurement of a source for each pixel $i$, we use a weighted average
of all measurements of that source falling with a single pixel.  The weights
are purely geometric and come from interpolating the measurement between the four
nearest pixels.

\subsection{Forming a Least-Squares Problem}
\label{lstsqs}

Once all the data are gridded, we solve Equation \ref{healmeas} for all
crossing pixels simultaneously.  To do this, we set up a linearized least-squares
problem using logarithms:   
\begin{equation}
\label{logmeas}
\log(M_i) = \log(b_i) + \log(f_k).
\end{equation}
Because thermal noise in measurements becomes non-Gaussian in Equation \ref{logmeas}, the
solution to the logarithmic formulation of the least-squares problem is biased.  
In \S\ref{beamsims}, we investigate the effect of this
bias using simulations and find that the
accuracy of our results is not limited by this bias, but instead by sidelobes of other
sources.  Therefore, while Equation \ref{basicmeas} can in principle be solved
without resorting to logarithms using an iterative least-squares approach, we
find that this is not necessary given our other systematics.  

Once the logarithms are taken, we can construct a solvable matrix equation, which can be expressed generally as:
\begin{equation}
\label{matrix2}
\mathbf{WM = WAX},
\end{equation}
where $\mathbf{W}$ is a column-matrix of weights, $\mathbf{M}$ is a column-matrix of measurements, $\mathbf{A}$ is the matrix-of-condition, describing which measurements are being used to constrain which parameters, and $\mathbf{X}$ is a column matrix containing all the parameters we wish to measure: the beam responses $b$ and the source flux densities $f$.
To
illustrate the form of this equation, we present the matrix representation of
the Cygnus A/Virgo A system shown in Figure \ref{trackexample}:
\begin{widetext}
\begin{align}
\label{matrix1}
{\left(\begin{array}{c}
W_1 \\
W_2 \\
W_3 \\
W_4 \\
W_5 \\
W_6 \\
W_7 \\
W_8 \\
W_9 \\
W_{10} \\
W_{11} \\
W_{12}
\end{array} \right)}
{\left(\begin{array}{c}
\log M_1 \\
\log M_2 \\
\log M_3 \\
\log M_4 \\
\log M_5 \\
\log M_6 \\
\log M_7 \\
\log M_8 \\
\log M_9 \\
\log M_{10} \\
\log M_{11} \\
\log M_{12}
\end{array} \right)}
=
{\left(\begin{array}{c}
W_1 \\
W_2 \\
W_3 \\
W_4 \\
W_5 \\
W_6 \\
W_7 \\
W_8 \\
W_9 \\
W_{10} \\
W_{11} \\
W_{12}
\end{array} \right)}
{\left( \begin{array}{cccccc|cc}
1 & 0 & 0 & 0 & 0 & 0 & 1 & 0 \\
1 & 0 & 0 & 0 & 0 & 0 & 0 & 1 \\
0 & 1 & 0 & 0 & 0 & 0 & 1 & 0 \\
0 & 1 & 0 & 0 & 0 & 0 & 0 & 1 \\
\hline \
0 & 0 & 1 & 0 & 0 & 0 & 1 & 0 \\
0 & 0 & 1 & 0 & 0 & 0 & 0 & 1 \\
0 & 0 & 0 & 1 & 0 & 0 & 1 & 0 \\
0 & 0 & 0 & 1 & 0 & 0 & 0 & 1 \\
\hline \
0 & 0 & 0 & 0 & 1 & 0 & 0 & 1 \\
0 & 0 & 0 & 0 & 1 & 0 & 0 & 1 \\
0 & 0 & 0 & 0 & 0 & 1 & 0 & 1 \\
0 & 0 & 0 & 0 & 0 & 1 & 0 & 1
\end{array} \right)}
{\left(\begin{array}{c}
\log b_1 \\
\log b_2 \\
\log b_3 \\
\log b_4 \\
\log b_5 \\
\log b_6 \\
\log f_{cyg} \\
\log f_{vir}
\end{array} \right)}
\end{align}
\end{widetext}

On the left-hand side of the equation are the two column matrices $\mathbf{W}$ and $\mathbf{M}$.  The weights, $W_i$, are defined in \S\ref{weighting}. 
$\mathbf{M}$ contains logarithms of the
perceived source flux density measurements in each pixel, $M_i$.  Recall that each $M_i$ corresponds to one source.  

On the right-hand side, the weighting column matrix $\mathbf{W}$ appears again, followed by 
the matrix-of-condition $\mathbf{A}$, and then $\mathbf{X}$, a 
column matrix containing the parameters we wish to solve for: the primary beam
response at the 6 crossing points and the flux densities of Virgo and Cygnus.
The matrix-of-condition, $\mathbf{A}$, identifies which
sources and crossing points are relevant for each equation.  We have
schematically divided it: to the left of the vertical line are the indices used
for selecting a particular crossing point; to the right are those for the
sources.  The first 4 lines represent the two northern Cygnus/Virgo crossing
points, and the next four represent the two southern ones (which are identical
copies of the northern ones).  Finally, the last 4 lines represent the 2 points
where Virgo crosses itself; notice that the Cygnus source column is blank for
these 4 rows.

It should be noted that Equations \ref{matrix2} and \ref{matrix1} contains no absolute flux density reference.  The simplest way to set this scale is to treat all the recovered flux densities as relative values compared to the flux calibrator (Cygnus A in the case presented here).  One can then place all the flux densities onto this absolute scale.  An equally valid approach is to append an extra equation with a very high weight, which sets the flux calibrator to its catalog flux value.

\subsubsection{Weighting of Measurements in the Least-Squares Formalism}
\label{weighting}

In a least-squares approach, optimal
weights are inversely proportional to the variance of each measurement.  To calculate
the variance of each measurement, we must propagate the uncertainty in the initial perceived flux density
measurements through the averaging and logarithm steps.
The noise level in each interferometric visibility is roughly constant, and  
the DDR filters average over a fixed number of visibilities for each perceived flux density estimate, leading to
equal variance at each time sample.
To produce optimal weights accounting for the many time
samples averaged into each beam pixel and the propagation of noise through the logarithm,
each
logarithmic measurement in Equation \ref{matrix2} should be weighted by:
\begin{equation}
W_i = \sqrt{M_i^2\sum\limits_j w_j},
\end{equation} 
where $j$ indexes the time step (which is generally fast enough to produce many measurements inside a pixel), $w_j$ is
the geometric sub-pixel interpolation weighting of each measurement, and $M_i$
is the weighted average of all measurements of a source's flux density in 
pixel $i$.  
Without the square root, these weights would be proportional to the inverse of the variance in each
logarithmic measurement; the sum over the geometric weights is the standard
reduction of variance for a weighted average, and the factor of $M_i^2$ comes
from propagating variances through the logarithm.  The square root appears because a least-squares solver using matrix inversion will add in an additional factor of the weight, leading to the desired inverse-variance weighting.  Other solvers using different methods may require different weights.

\subsubsection{Solving the Least-Squares Equation for a First-Order Beam Model}

\begin{figure}
\centering
\includegraphics[scale=.4]{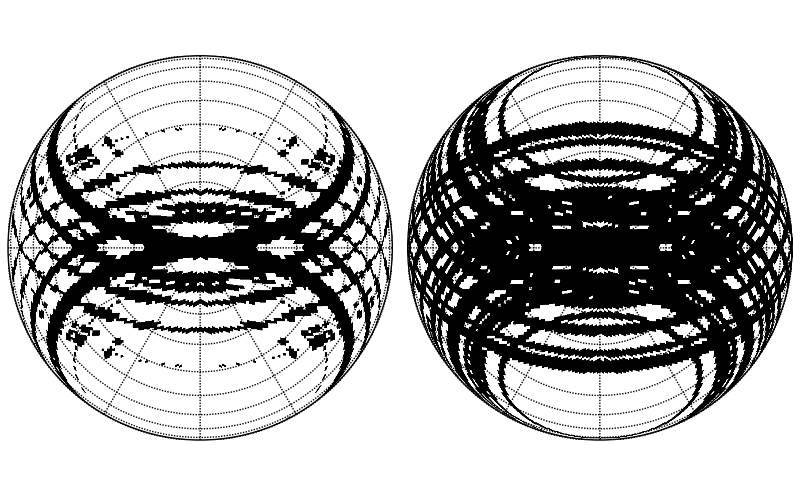}
\caption{
The sky coverage of crossing points and source tracks.  \emph{Left:} The location of crossing points for the 25 sources listed in Table \ref{srctable} used to calibrate the PAPER primary beam.  The projection is orthographic with zenith at the center; dotted lines are
$10^{\circ}$ and $30^{\circ}$ steps in elevation and azimuth, respectively.
\emph{Right:} The sky coverage of all 25 source tracks.  Although the least-squares inversion solves only for the beam response at crossing points, we can include all source track data using the recovered flux density of each source to create a primary beam estimate along the entire track.
}
\label{cross_v_track}
\end{figure}

The solution of Equation \ref{matrix2}, the matrix $\mathbf{A}$, contains two distinct sets of parameters:
the beam responses at crossing points and the flux densities of each source (modulo an absolute scale).  
There is additional information that may be included to improve the
beam model beyond that generated by solving for the responses at crossing points.
Given the flux-density solutions for each source, each source track now provides
constraints on beam pixels that are not crossing points.  By dividing each
perceived flux density source track by the estimated inherent flux density from
the least-squares inversion, we produce a track of primary beam responses, with
greater coverage than that
provided by crossing points alone.  
We illustrate the difference in coverage in Figure \ref{cross_v_track}.  
The left hand panel shows the locations of crossing points for the 25 sources used to calibrate the PAPER beam, listed in Table \ref{srctable}.  The right hand panel shows the increased beam coverage that comes from including non-crossing point source-track data.
To create an initial beam model, we average within each pixel the estimated
beam responses from each source, weighting by the estimated flux of that
source.
Given equal variance in each initial perceived flux density measurement, this choice of weights is weighting by signal to noise.

\subsection{Using Deconvolution to Fill in Gaps in the Beam Model}
\label{deconvolution}

To produce a model of the beam response in any direction, we must fill in the
gaps left by limited sky sampling.  We use a CLEAN-like \citep{hogbom1974}
deconvolution algorithm in Fourier space to fill in the holes in the beam.  We
iteratively fit and remove a fraction of the brightest Fourier component of our
beam until the residuals between the model and the data are below a specified
threshold.  In addition to measured beam responses derived from source tracks, we add the constraint
that the beam must go to zero beyond the horizon.  In the deconvolution, each pixel
is weighted by the estimate of the beam response in that pixel, reflecting that SNR will
be highest where the beam response is largest.  This beam-response weighting was unnecessary
in previous steps, since the least-squares approach solved for each
pixel independently.  This weighting scheme again represents signal-to-noise weighting, given the equal variance in our initial perceived flux density estimates.  The result of the deconvolution is an interpolated primary
beam model with complete coverage across the sky.  

\subsection{Introduction of Prior Knowledge}
\label{priors}

Up to this point, we have made only two fairly weak assumptions about our
beam: that it possesses $180^{\circ}$ rotational symmetry, and that the response
is zero below the horizon.  However, if we have additional prior knowledge
about our beam, we can better constrain the final model.  In particular, we can
use beam smoothness constraints to identify unphysically small scale features
introduced by sidelobes in the source extraction or by incomplete sampling in the deconvolution.  
We choose to 
incorporate additional smoothness information by filtering our model in
Fourier space to favor large-scale modes.  

PAPER dipoles were designed with emphasis on spatial and
spectral smoothness in primary beam shape.  To smooth the PAPER beam model, we choose
a cutoff in Fourier space that corresponds to the scale at which $>99.9\%$ of
the power is accounted for in a computed electromagnetic model of the
beam.  While such a filter is not necessarily generalizable to other antenna
elements, we find it necessary to suppress the substantial sidelobes associated
with observations from a 12-antenna PAPER array that are discussed below.

\section{Application to Simulated Data}
\label{sims}

To test the robustness of this approach, we apply it to several simulated
data sets.  The results of these simulations, with and without Gaussian noise,
are described below in \S\ref{beamsims}.  We also simulate raw visibility data
to test the effectiveness of source extraction; these simulations are discussed
in \S\ref{vissims}.  The major difference between these two methods of
simulation is that the simulation using raw visibility data allows for imperfect source isolation, leading to
contamination of the source tracks by sidelobes of other sources.  These
sidelobes have a significant effect on the final beam model that is derived.

\subsection{Simulations of Perceived Flux Density Tracks}
\label{beamsims}

We simulate perceived flux density tracks using several model beams of
different shapes, including ones with substantial ellipticity and an
$\sim15^{\circ}$ rotation around zenith.  We also input several source
catalogs, including a case with $10,000$ Jy sources spaced every degree in
declination, and a case using the catalog values and sources listed in Table
\ref{srctable} that approximately match the sources extracted from
observations.  In all combinations of beam models and source catalogs, we
recover the input beam and source values with $<5\%$ error.  The average
error in source flux density is 2.5\%.  We see no evidence for residual
bias, as the distribution of error is consistent with zero mean to within one
standard deviation.

We also test the effect of adding various levels of Gaussian noise to a
simulation involving a fiducial beam model and the 25 sources listed in Table
\ref{srctable}.  Only when the noise level exceeds an RMS of 10 Jy in each perceived flux density measurement
does the mean error in the solutions exceed 10\%.  As the
noise increases beyond this level, there is a general trend to bias recovered
flux densities upward; as mentioned earlier, this bias is introduced by the
logarithms in Equation \ref{logmeas}.  However, the expected corresponding
noise level of DDR-extracted perceived flux density measurements from our 12-element PAPER array is $<1$ Jy per sample.  
At a
simulated rms noise level of 1 Jy, the solutions are recovered with a mean error of
$<3\%$.  This result validates our previous statement that the bias
introduced by the logarithms in Equation \ref{logmeas} is not a dominant source
of error.

It is also worth noting that these simulations were used to identify $0.9^{\circ}$ as the best
HEALPix pixel side for our grid; with too large a pixel size ($1.8^{\circ}$)
the
model becomes significantly compromised.
This results from combining measurements of sources that are subject to
significantly different primary beam responses.  We choose not to use a smaller pixel size, since it will reduce the fractional sky coverage of our crossing points and increase the computational demand of the algorithm.  

\subsection{Simulations of Visibilities}
\label{vissims}

We also apply this technique to simulated visibilities 
in order to
test the complete analysis pipeline, including the DDR-based estimation of
perceived source flux densities.  The visibility simulations are implemented in
the AIPY\footnote{http://pypi.python.org/pypi/aipy/} software toolkit.  The
simulated observations correspond to actual observations made with the 12-element
PAPER deployment in Green Bank described in \S\ref{data}.  We match the
observations in time, antenna position, and bandwidth.  We also include the
expected level of thermal noise in the simulated visibilities.  We simulate
``perceived" visibilities by attenuating the flux density of each source by a model
primary beam.  The DDR filters return estimates of perceived source flux density
for the input primary beam.

In these simulations, we only include bright point sources and a uniform-disk model
of the Sun.
As a result, source extraction is expected to be more
accurate in simulation than in real data, since the sources we
extract account for $100\%$ of the simulated signal.  However, these
simulations do provide a useful test of DDR-filter-based source extraction and
of the level of contamination
from sidelobes.

\begin{figure}
\centering
\includegraphics[scale=.45]{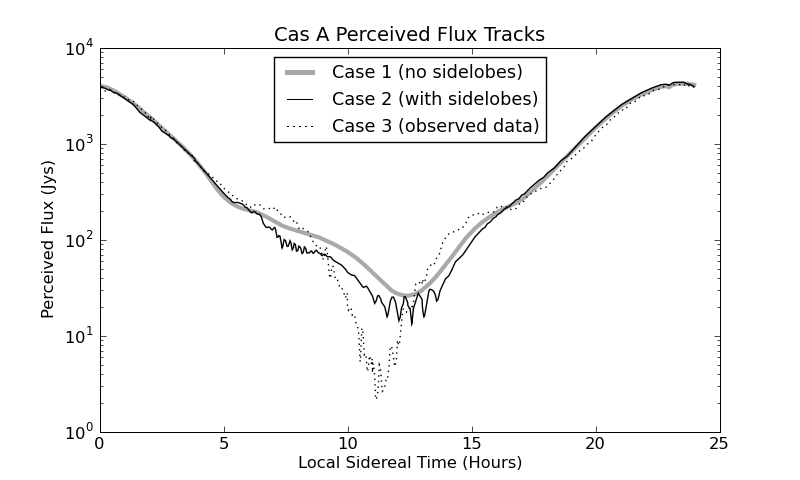}
\caption{
Three different perceived flux tracks for Cassiopeia A.  Case 1 is simply a cut
through a model beam multiplied by a catalog flux density value; this is the
simplest case and contains no sidelobe contamination.  Case 2 uses DDR filters
to extract the source track from simulated visibilities; case 3 does the same,
but applied to observed data.  The simulated visibilities in case 2 demonstrate the
features that arise from sidelobes of other sources during the DDR source
extraction.  However, the simulation employs a simplistic point-source sky and
cannot reproduce all sidelobe features seen in the real data.
}
\label{trackcompare}
\end{figure}

As might be expected, the estimates of perceived source flux density versus
time from simulated visibilities contain structure that is not attributable to
the beam.  These features are almost all due to sidelobes of other sources;
they persist in real data and are reproducible over many days.  Figure
\ref{trackcompare} shows the difference in the source tracks produced for
Cassiopeia A by each simulation method (cases 1 and 2 in the figure), as well
as the track extracted from the observed data described \S\ref{data} (case 3).  
We
find that the DDR filter only extracts a fraction of the flux density from
the Crab Nebula.  This bias is unique to Crab, and is most likely a result
of the proximity of the Sun to Crab in these observations, coupled with
the limited number of independent baselines in the 12-antenna array. 
For this reason, we choose to exclude the Crab from our final analysis. 

Performing the least-squares inversion described in \S\ref{lstsqs} on the
simulated source tracks, we find that the sidelobes introduce errors into
estimates of source flux densities.  The average error in flux density is 10\%,
with the largest errors exceeding 20\%.   The distribution of errors is
consistent with zero mean, indicating no strong biasing of the flux densities.

\begin{figure}
\centering
\includegraphics[scale=.45]{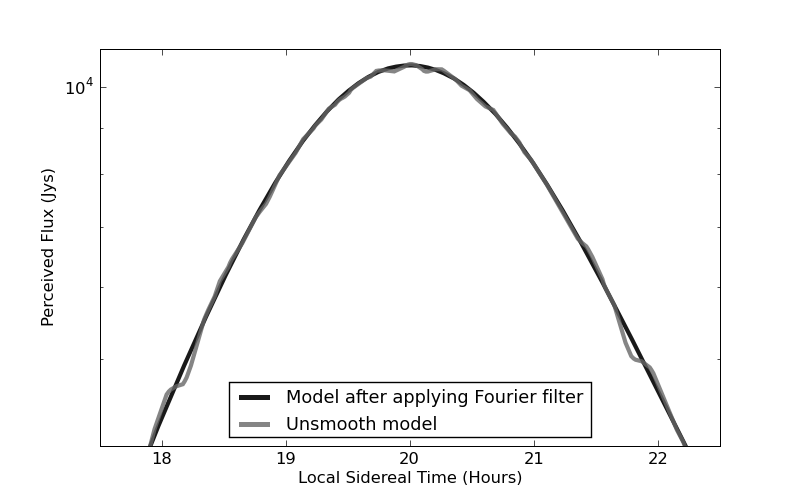}
\caption{
The effect of the Fourier domain filter.  The deconvolution of
\S\ref{deconvolution} fills in gaps in sky coverage, but leaves 
small scale structure that is not associated with the primary beam (gray curve).  
By filtering the solution in the Fourier
domain to incorporate prior knowledge of beam smoothness, one can achieve an improved beam model
(black curve).
}
\label{filter}
\end{figure}

We also find that the resultant primary beam model matches the input beam to
within 15\% percent.  However, the model contains small scale variations, as
seen in Figure \ref{filter}.  
Using prior knowledge of beam smoothness,
we can reject these features as unphysical and apply a Fourier-domain filter, as
described in \S\ref{priors}.  This filter reduces the effect of
sidelobes in the source tracks, as they appear on Fourier scales that are not
allowed by smoothness constraints.  Even with a relatively weak prior on
smoothness (i.e., retaining more Fourier modes than are present in our input
model), we substantially reduce the small scale variations in our beam.  The
effect of the filter on small scale variations is illustrated in Figure
\ref{filter}.  The filter also reduces large scale errors in the model,
bringing the overall agreement to with 10\% of the input beam.

In summary, we find that although interfering sidelobes from other sources
compromise our source flux density measurements, our approach recovers the
input primary beam model at the 15\% level.  With the introduction of a Fourier
space filter motivated by prior knowledge of the beam smoothness, the output model
improves to within 10\% of the input.  It is also worth noting that the
effectiveness of the beam calibration technique presented here will substantially improve
with larger arrays, where sidelobe interference will be reduced and source
tracks will more closely resemble the ideal case discussed in \S\ref{beamsims}.  

\section{Observed Data}
\label{data}

\begin{figure}
\centering
\includegraphics[scale=.6, trim=2cm 0cm 2cm 0cm, clip=true]{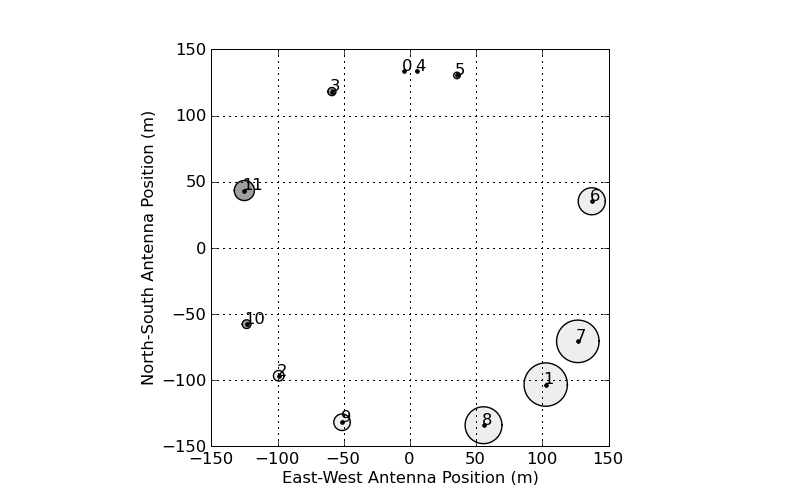}
\caption{
The 12 element PAPER array in Green Bank.  Shaded circles represent vertical height relative to antenna 0; the radius of the circle is 10 times the height in meters.  Light gray is up, dark gray is down.
}
\label{antpos}
\end{figure}

In this section, we describe the application of this technique to 24 continuous
hours of data taken with the PAPER array deployed at the NRAO site near Green Bank from July 2 to July
3, 2009.  At this time, the array consisted of 12 crossed-dipole elements.  Only the
north/south linear polarization from each dipole was correlated.  The data used in this analysis were observed
between 123 to 170 MHz (avoiding RFI) and were split in 420 channels.
The array was configured in a ring of 300m radius.  The longest baselines are
300m, while the shortest baseline is 10m; the configuration is shown in
Figure \ref{antpos}.  This configuration gives
an effective image plane resolution of $0.4^{\circ}$.

\subsection{Data Reduction}

In addition to the DDR-filter algorithm described in \S\ref{pfluxes}, several
other pre-processing steps are necessary.  Per-antenna phase and amplitude
calibration are performed by fringe fitting to Cygnus A and other bright
calibrator sources.  We also use a bandpass calibration based on the spectrum
of Cygnus A.  This calibration is assumed to be stable over the full 24 hours.
Two further steps in the reduction pipeline were first described in
\citet{parsons_et_al2010}: gain linearization to mitigate data quantization
effects in the correlator, and RFI excision.  

\begin{figure}
\centering
\includegraphics[scale=.45]{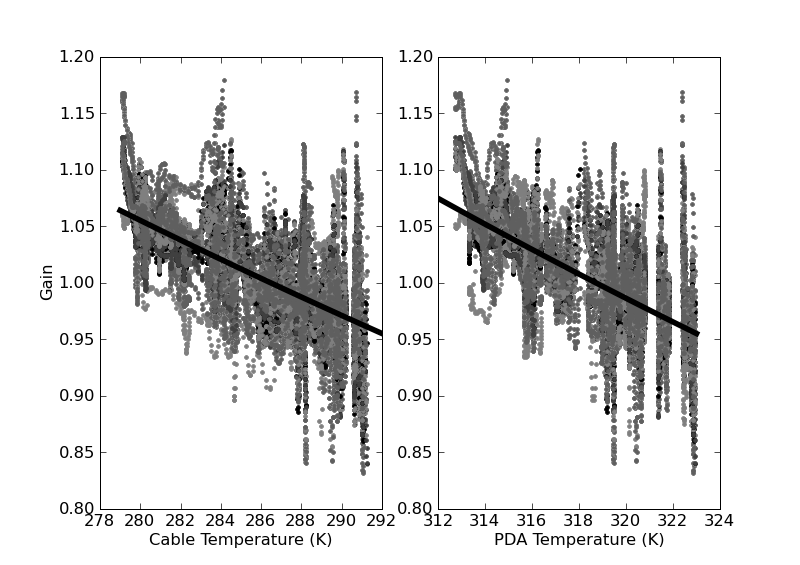}
\caption{
The effect of ambient temperature on the system gain, derived from
observations of Cygnus A near zenith, where beam effects are unimportant.  The
left-hand side shows the effect of the cable temperature; the right shows
pseudo-differential amplifier (PDA)/balun temperature.  The solid black line is
the best fit to the data.  Different shades of gray represent data from
different antennas.
}
\label{tdepgains}
\end{figure}

New to this work is the use of the cable and balun temperatures to remove
time-dependent gains caused by temperature fluctuations.  
A thirteenth dipole was operated as the ``gain-o-meter" described in
\citet{parsons_et_al2010}.  In brief, the ``gain-o-meter" is an antenna where the
balun is terminated on a matched load, rather than a dipole.  The measured noise power
from this load tracks gain fluctuations in the system.  
We record the temperature of several
components of our ``gain-o-meter" using the system described in
\citet{parashare_and_bradley2009}.
We find a strong correlation between the absolute gain of
our system and the temperature of these components; this effect is illustrated
in Figure \ref{tdepgains}.  
We correct for these gain changes by applying a
linear correction derived from the measured temperature values.  This step is crucial
for the success of the beam calibration; without correcting for them, these
gain drifts are indistinguishable from an east/west asymmetry in the primary
beam.  After these steps, we process the data with the DDR-filtering algorithm
to produce estimates of perceived flux density versus time for the 25 sources
listed in Table \ref{srctable}.

\subsection{Results}

\subsubsection{PAPER Primary Beam Model}

\begin{figure}
\centering
\includegraphics[scale=0.62, trim=3.7cm 0cm 2cm 0cm, clip=true]{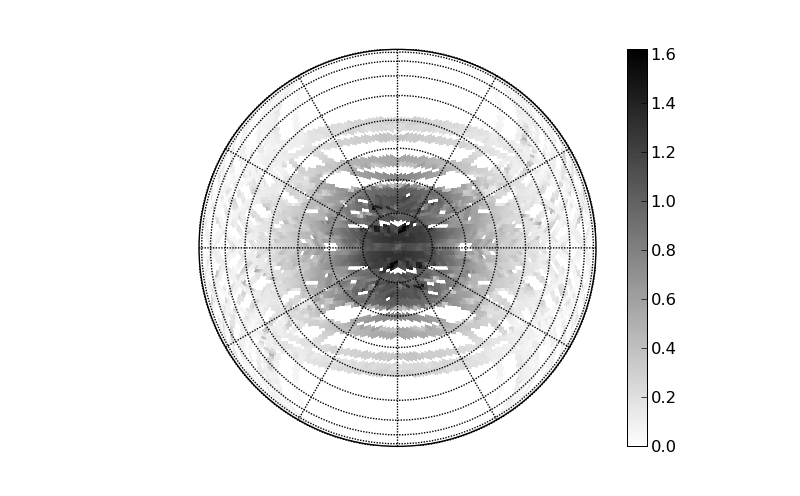}
\caption{
The initial solution for the PAPER primary beam response along source tracks.
Each perceived flux density track is divided by the estimate of that source's
inherent flux density produced by the least-squares inversion.  This yields
estimates of the primary beam response along each track.  With this initial
solution, a large fraction of the sky is already covered by the data, although
there is a fair bit of variation from point to point.  The color scale is linear
and normalized to 1.0 at zenith; as in Figure \ref{trackexample}, dotted lines
are $10^{\circ}$ and $30^{\circ}$ steps in elevation and azimuth, respectively.
}
\label{dirtybeam}
\end{figure}

The first estimate of the PAPER primary beam derived with this approach is
shown in Figure \ref{dirtybeam}.  
This figure is produced by dividing each
source track by an estimate of its inherent flux density produced by the least
squares inversion.  This transforms each track into an estimate of the primary beam
response, which are the added together into a HEALPix map.  There are
significant fluctuations from pixel to pixel that are clearly unphysical, and relate to source sidelobes.
There are also significant gaps in beam coverage, even with 25 source tracks.

\begin{figure}
\centering
\includegraphics[scale=0.62, trim=3.7cm 0cm 2cm 0cm, clip=true]{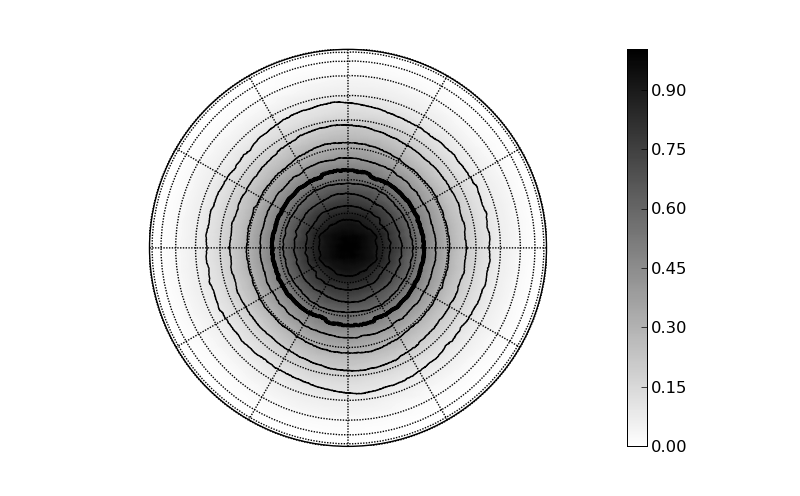}
\caption{
The deconvolved PAPER beam model.  The grayscale colors are linear and show the
primary beam response, normalized to 1.0 at zenith.  Solid black lines are
contours of constant beam response in 10\% increments; the thick black line is the half-power point.  As in Figure
\ref{trackexample}, dotted lines are $10^{\circ}$ and $30^{\circ}$ steps in
elevation and azimuth, respectively.  This model was produced by deconvolving
the sampling pattern from Figure \ref{dirtybeam}, interpolating over 
the gaps in declination where there are no strong calibrator sources.
}
\label{cleanbeam}
\end{figure}

\begin{figure}
\centering
\includegraphics[scale=0.62,trim=3.7cm 0cm 2cm 0cm, clip=true]{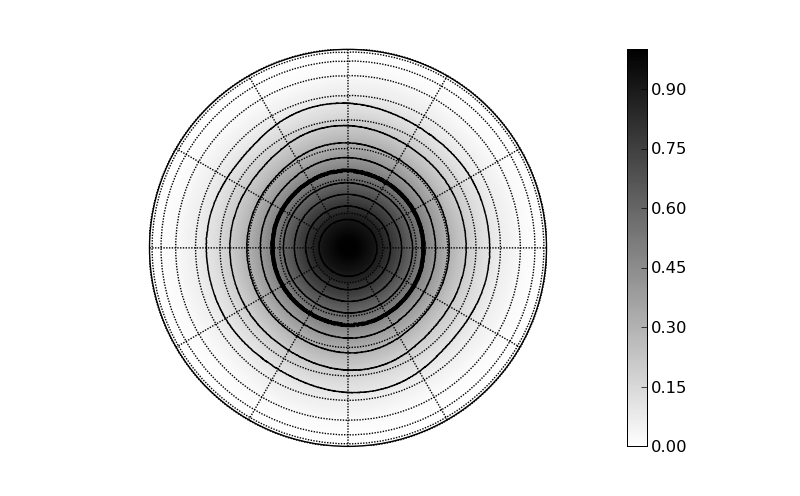}
\caption{
The smooth PAPER beam model.  As in Figure \ref{cleanbeam}, the grayscale
colors are linear and show the primary beam response, normalized to 1.0 at
zenith.  Solid black lines are contours of constant beam response in 10\%
increments; the thick black line is the half-power point.  Dotted lines are $10^{\circ}$ and $30^{\circ}$ steps in elevation
and azimuth, respectively.  This smooth model was produced by retaining only the large scale
Fourier components from Figure \ref{cleanbeam}.  The cutoff in Fourier space
was derived from a computed electromagnetic PAPER beam simulation.  This
Fourier mode selection substantially smooths out variations in the initial
solution that are unphysical based on the smoothness of beam
response in the simulation.
}
\label{filterbeam}
\end{figure}

Figure \ref{cleanbeam} show the results after deconvolving the sampling pattern
from Figure \ref{dirtybeam}.  
The beam is now complete across the entire sky,
although there is still substantial small-scale structure unrelated to the inherent beam response.  We suppress these features in our final model by retaining only
the large scale Fourier components, as described in \S\ref{deconvolution}.
This final model is shown in Figure \ref{filterbeam}.
It is clear that with our
deconvolution and choice of Fourier modes, we recover a very smooth, slowly
evolving beam pattern, as expected from previous models of the PAPER primary beam.

\subsubsection{Source Catalog}

The measured flux densities of the calibrator sources, produced by the
least-squares inversion, are presented in Table \ref{srctable}.  The overall
flux scale is normalized to the value of Cygnus A reported in
\citet{baars_et_al1977}.  Error-bars are estimated by measuring the change
in flux necessary to increase $\Delta\chi^2$ between the model and the measured
perceived source flux density tracks by 1.0.  As noted in
\S\ref{vissims}, these flux densities can be compromised by sidelobes of
other sources.  However, the results are generally accurate to within $\pm10\%$.

\subsection{Tests of Validity}

We perform several tests to investigate the validity of our results.  The
comparison of our recovered fluxes to their catalog values shows generally good
agreement.  As argued in \citet{vollmer_et_al2005} and \citet{jacobs_et_al2011},
there is considerable difficulty in accurately comparing catalogs, particularly
at these frequencies.  Differences in interferometer resolution and synthesized
beam patterns can lead to non-trivial disagreement in measured source flux densities.
Given this fact, and the complicated sidelobes associated with a 12-element array
present in our method, we do not find the lack of better agreement with catalog
fluxes troubling.

To test the stability of our pipeline, we test a separate observation spanning
the following day.  The beam model produced by this data matches our first
model to within 2.5\%, and source flux densities remain consistent to 5\%.
This test confirms our suspicion that our errors are not dominated by random
noise.

\section{Conclusions}
\label{conclusions}

We have presented a new technique for calibrating the primary beam of a
wide-field, drift-scanning antenna element.  The key inputs to the method are
measurements of perceived flux density of individual sources versus time as
they drift through the beam.  In this paper, we use delay/delay-rate filters
\citep{parsons_and_backer2009} as estimators in a self-calibration loop
to extract such measurements from raw
interferometric data.  However, the remainder of our method is agnostic to how
these tracks are derived.  The only assumption necessary to make this an
over-constrained and solveable problem is $180^{\circ}$ rotational
symmetry in the beam.  With this assumption, we create ``crossing points" where
there is enough information for a least-squares inversion to solve for both the
inherent flux density of each source and the response of primary beam at the
crossing points.

We test this approach using simulated tracks of perceived source flux density
across the sky and simulated visibilities.  Using the source tracks, the
least-squares inversion reliably recovers the primary beam values and source
flux densities in the presence of noise 10 times that present in a 12-element
PAPER array.  The simulated visibilities demonstrate that sidelobes of other
bright sources are the source of the dominant errors in source extraction when
only 12 antennas are used.  The presence of these features in the perceived
source flux densities limits the accuracy of estimates of inherent source flux
densities.  However, in simulation, we are able to recover a primary beam model
accurate to within 15\% percent and source flux densities with an average error
of 10\%.  Using prior information regarding beam smoothness, we improve our
model to better than 10\% accuracy.

While these caveats about the effectiveness of the least-squares technique in
the presence of sidelobes may seem worrisome, it bears repeating that these are
issues with data quality and not with the technique itself.  For example, data
from a 32-element PAPER array has shown that DDR filters can extract the
sources used in this analysis with little systematic biases.  (We do not use
this data here do the lack of a temperature record for gain stabilization and
the presence of a particularly active Sun when the array was operating.)
Therefore, it seems that this technique has significant potential for precise
beam calibration on larger arrays.

Another future goal for this technique is to calibrate the frequency-dependence
of the beam.  Here, we have used our entire bandwidth to improve signal-to-noise in source
extractions.  For a larger array with higher SNR and lower sidelobes, our
perceived source flux density measurements can be cut into sub-bands to look at
the beam as a function of frequency.

Finally, it is possible to forgo the assumption of $180^{\circ}$ rotational
symmetry altogether and allow for possible north/south variation in the beam.
An experiment in which the dipoles are \emph{physically} rotated on a daily
basis can be used to create the same kind of crossing points, since one has
changed the section of the beam each source crosses through.  Work is
progressing on such an experiment using the PAPER array.

\section{Acknowledgements}

The PAPER project is supported through the NSF-AST program (award \#0804508), and
by significant efforts by staff at NRAO's Green Bank and Charlottesville sites.
PAPER acknowledges the significant correlator development efforts of Jason R. Manley.  
We also thank our reviewer for their helpful comments.
ARP acknowledges support from the NSF Astronomy and Astrophysics Postdoctoral Fellowship under 
award AST-0901961.

\newpage
\ctable[caption=Measured flux densities for all the sources used in this work.  Catalog values come from the 3C catalog \citep{edge_et_al1959} at 159 MHz unless otherwise noted.  Errors are estimated by measuring the change in flux required to increase $\Delta\chi^2$ between the model and measured perceived flux density source tracks by 1.0., width=\textwidth, label=srctable]{c|c||c|c|c}%
{\tnote[a]{3CRR \citep{laing_et_al1983} at 178 MHz.}
\tnote[b]{Culgoora \citep{slee1995} at 160 MHz.}
\tnote[c]{PSA32 \citep{jacobs_et_al2011}.  Measurements made by PAPER.}
\tnote[d]{Extended (35x27 arcmin) supernova remnant (W44).}
\tnote[e]{Flux density calibrator, using values from \citet{baars_et_al1977}.} 
\tnote[f]{Extrapolated using the formula in \citet{baars_et_al1977}.  This is probably a substantial underestimate of the flux density, as suggested by \citet{helmboldt_and_kassim2009}.}
}
{RA & Dec & Measured Flux Density (Jy) & Catalog Flux Density (Jy) & Name \\
\hline \hline
1:08:54.37 & +13:19:28.8 & $78.1^{+14.7}_{-24.4}$ & 58, 49\tmark[a] & 3C 33 \\
1:36:19.69 & +20:58:54.8 & $47.3^{+12.3}_{-37.3}$ & 27 & 3C 47 \\
1:37:22.97 & +33:09:10.4 & $60.0^{+10.2}_{-13.7}$ & 50 & 3C 48 \\
1:57:25.31 & +28:53:10.6 & $36.2^{+9.8}_{-23.6}$ & 7.5, 16\tmark[b], 23\tmark[c] & 3C 55 \\
3:19:41.25 & +41:30:38.7 & $81.7^{+12.0}_{-16.5}$ & 50 & 3C 84 \\
4:18:02.81 & +38:00:58.6 & $75.9^{12.2}_{-14.0}$ & 60 & 3C 111 \\
4:37:01.87 & +29:44:30.8 & $205.3^{+26.9}_{-32.1}$ & 204 & 3C 123 \\
5:04:48.28 & +38:06:39.8 & $100.3^{+17.0}_{-20.2}$ & 85 & 3C 134 \\
5:42:50.23 & +49:53:49.1 & $56.5^{+9.7}_{-11.5}$ & 63 & 3C 147 \\
8:13:17.32 & +48:14:20.5 & $66.3^{+9.5}_{-9.9}$ & 66 & 3C 196 \\
9:21:18.65 & +45:41:07.2 & $48.5^{+7.5}_{-9.0}$ & 42 & 3C 219 \\
10:01:31.41 & +28:48:04.0 & $40.3^{+8.7}_{-15.6}$ & 30 & 3C 234 \\
11:14:38.91 & +40:37:12.7 & $34.5^{+7.0}_{-19.9}$ & 21.5 & 3C 254 \\
12:30:49.40 & +12:23:28.0 & $1108.9^{+29.6}_{-28.1}$ & 1100 & Vir A \\
14:11:21.08 & +52:07:34.8 & $64.4^{+6.4}_{-6.7}$ & 74 & 3C 295 \\
15:04:55.31 & +26:01:38.9 & $73.5^{+7.4}_{-9.0}$ & 72 & 3C 310 \\
16:28:35.62 & +39:32:51.3 & $60.5^{+7.2}_{-12.4}$ & 49 & 3C 338 \\
16:51:05.63 & +5:00:17.4 & $461.7^{+78.6}_{-112.7}$ & 325\tmark[a], 378\tmark[b], 373\tmark[c] & Her A \\
17:20:37.50 & -0:58:11.6 & $228.8^{+61.7}_{-147.5}$ & 180, 236\tmark[a], 276\tmark[b], 215\tmark[c] & 3C 353 \\
18:56:36.10 & +1:20:34.8 & $251.3^{+60.0}_{-135.8}$ & 680 & 3C 392\tmark[d] \\
19:59:28.30 & +40:44:02.0 & $10622.9^{+126.0}_{-128.9}$ & 10623 & Cyg A\tmark[e] \\
20:19:55.31 & +29:44:30.8 & $62.2^{+16.4}_{-28.0}$ & 36 & 3C 410 \\
21:55:53.91 & +37:55:17.9 & $51.9^{+11.5}_{-16.2}$ & 43 & 3C 438 \\
22:45:49.22 & +39:38:39.8 & $71.3^{+12.4}_{-19.3}$ & 50 & 3C 452 \\
23:23:27.94 & +58:48:42.4 & $9198.1^{+203.4}_{-186.9}$ & 6230\tmark[f] & Cas A
}

\end{document}